# Quantum Poisson Solver without Arithmetic

Shengbin Wang[1,†], Zhimin Wang[1,†], Guolong Cui[1], Shangshang Shi[1], Ruimin Shang[1], Jiayun Li[1], Wendong Li[1], Zhiqiang Wei[2,3,*] and Yongjian Gu[1,*]

[1] College of Physics and Optoelectronic Engineering, Ocean University of China, Qingdao 266100, China

[2] College of Computer Science and Technology, Ocean University of China, Qingdao 266100, China

[3] High Performance Computing Center, Pilot National Laboratory for Marine Science and Technology (Qingdao), Qingdao 266100, China

[†] These authors contributed equally: Shengbin Wang, Zhimin Wang

[*] Author to whom any correspondence should be addressed.

E-mail: yjgu@ouc.edu.cn; weizhiqiang@ouc.edu.cn

**Abstract**

Solving differential equations is one of the most promising applications of quantum computing. Most existing quantum algorithms addressing general ordinary and partial differential equations are thought to be too expensive to execute successfully on Noisy Intermediate-Scale Quantum (NISQ) devices. Here we propose an efficient quantum algorithm for solving one-dimensional Poisson equation based on the simple $R_y$ rotations. Our quantum Poisson solver (QPS) avoids the need to perform the expensive routines such as phase estimation, quantum arithmetic or Hamiltonian simulation, etc. The complexity of our QPS is $3n$ in qubits and $5/3n^3$ in one- and two-qubit gates, where $n$ is the logarithmic of the number of discrete points. An overwhelming reduction of constant factors of big-O complexity is achieved, which is critical to evaluating the practicality of actually implementing the algorithm on a quantum computer. In terms of the error $\varepsilon$, the complexity is $\log(1/\varepsilon)$ in qubits and $\mathrm{poly}(\log(1/\varepsilon))$ in operations. The algorithms are demonstrated using a quantum virtual computing system, and the circuits are executed successfully on the IBM real quantum computers. The present QPS may represent a potential real application for solving differential equations on NISQ devices.

## 1. Introduction

Quantum computing is capable of solving problems efficiently which are intractable for classical computing. Since the discovery of Shor's factoring algorithm [1], quantum algorithms demonstrating advantages over classical methods have been developed for a substantial variety of tasks [2]. One of the most promising applications is for solving differential equations which is the main task in high-performance scientific computing.

A series of quantum algorithms for solving ordinary and partial differential equations

(ODEs and PDEs) have been developed, which promise an exponential speed-up over classical algorithms in the dimension of equations and solution errors [3-9]. The main idea of these algorithms is that encoding the differential equations as a linear system, and solving it by Quantum Linear Systems Algorithm (QLSA) [10-12] or Hamiltonian simulation [13,14]. These algorithms are general methods for relatively general differential equations, but they are considered to be out of reach of large-scale fault-tolerant quantum computers.

In this paper, we focus on a relatively particular problem that is solving the one-dimensional Poisson equation with much lower cost. The Poisson equation, usually expressed as $\nabla^2\varphi = f$, is a widely used PDE across many areas of physics and engineering [15]. Many quantum algorithms are proposed to solve the *d*-dimensional Poisson equations [7,9,16-18]. Compared with them, the present quantum Poisson solver (QPS) has a much lower cost in qubits and quantum operations, and it aims to be implementable on Noisy Intermediate-Scale Quantum (NISQ) devices for real applications.

The inspiration of our method is from the following observation: the eigenvalues of the discretized matrix of the Poisson equation are square of sine values and sine values can be prepared as probability amplitudes easily by single-qubit $R_y$ rotation. The QLSA is actually to produce a quantum state with probability amplitudes being the reciprocals of eigenvalues. There should exist a simple way of accomplishing it by operating $R_y$ rotation only. That is, all the transformations are performed on the probability amplitude. With such a way, we could eliminate the need for phase estimation, Hamiltonian simulation [13,14] and quantum arithmetic [19-21], which are the major contributors to the complexity of the algorithm.

It is intriguing that the above inspiration can be realized naturally from the fact that the product of all eigenvalues of the discretized matrix equals constant. That is referred to as sine formula which is derived from the study of Cartan matrix in Lie algebra [22]. Combining the sine formula and single-qubit $R_y$ rotation, we establish our efficient QPS with a complexity of $3n$ in qubits and $5/3n^3$ in one- and two-qubit gates, where $n$ is the logarithmic of the dimension of the linear system. The present QPS is executed successfully on IBM near-term superconducting quantum computers and it would make our QPS the first rudiment of differential equation solver for NISQ devices. With the rapid developments of quantum hardware and techniques [23-26], our QPS is expected to be practical on NISQ devices.

## 2. The problem considered

We start with a high-level overview of the problem we address and the general idea of our method, before delving into details of the algorithm. The problem is solving the one-dimensional Poisson equation with Dirichlet boundary conditions, which can be described as

$$
\begin{aligned}
&-\frac{d^2v(x)}{dx^2}=b(x), x\in(0,1), \\
&v(0)=v(1)=0.
\end{aligned} \tag{1}
$$

The $b(x)$ is a given smooth function representing, say, velocity distribution in fluid dynamics problems, and $v(x)$ is the unknown to solve.

We choose the most simple central difference approximation to discretize the second-order derivative, then equation (1) turns to be

$$
\begin{aligned}
&h^{-2}(-v_{i-1}+2v_i-v_{i+1})=b_i+\varepsilon_i, i=1,2...,N-1, \\
&v_0=v_N=0.
\end{aligned} \tag{2}
$$

The number of discrete points is $N$+1 and the mesh size $h$ equals to $1/N$. The $\varepsilon_i$ represents the truncation error induced by the central difference approximation which is $O(h^2\cdot\|d^4v/dx^4\|_\infty)$ [27]. When $b(x)$ is smooth enough, the norm of the fourth derivative is bounded uniformly independent of $h$, and the error becomes $\varepsilon=O(h^2)=O(1/N^2)$. As shown below, this truncation error is the only error of the final solution state obtained by our method.

Ignoring the truncation error in equation (2), the problem of solving one-dimensional Poisson equation transfers into solving the following linear system of equations,

$$
h^{-2}\begin{pmatrix} 2 & -1 & & 0 \\ -1 & \ddots & \ddots & \\ & \ddots & \ddots & -1 \\ 0 & & -1 & 2 \end{pmatrix}\cdot\begin{pmatrix} v_1 \\ v_2 \\ \vdots \\ v_{N-1} \end{pmatrix}=\begin{pmatrix} b_1 \\ b_2 \\ \vdots \\ b_{N-1} \end{pmatrix} \Rightarrow A\vec{v}=\vec{b}. \tag{3}
$$

Matrix $A$ is a well-studied tridiagonal Toeplitz matrix. The eigenvectors and corresponding eigenvalues are $u_j(k)=\sqrt{2/N}\sin(j\pi k/N)$ and $\lambda_j=4N^2\sin^2(j\pi/2N)$, respectively [27]. Such a linear system of equations can be, of course, solved using the standard QLSA as done in refs. [16,18]. The solution state $|v\rangle$ would be expressed as

$$
|v\rangle=A^{-1}|b\rangle=(\sum\nolimits_j\frac{1}{\lambda_j}|u_j\rangle\langle u_j|)\cdot(\sum\nolimits_{j'}\beta_{j'}|u_{j'}\rangle)=\sum\nolimits_j C\frac{\beta_j}{\lambda_j}|u_j\rangle, \tag{4}
$$

where $C$ is a normalizing constant. In order to create such a state, the HHL algorithm first estimates eigenvalues $\lambda_j$ through phase estimation process, then calculates the reciprocals of $\lambda_j$ by arithmetic, and then transduce the reciprocals into the probability amplitudes through controlled rotation process [10].

Here we propose a much simple way to create the solution state based on the $R_y$ rotation. The fundamental observation is that matrix $A$ is a Cartan matrix, and there exists a beautiful sine formula as follows [22],

$$
2^{2^{n+1}-2}\prod\nolimits_{j=1}^{2^n-1}\sin^2(\frac{j}{2^{n+1}}\pi)=2^n, \tag{5}
$$

where $n$ is the logarithmic of the matrix dimension, namely $n = \log(N)$. Note that each term of sine square in the equation is actually the eigenvalue $\lambda_j$ of matrix $A$ up to a constant. Therefore, the reciprocal of an eigenvalue equals to the product of all the other eigenvalues.

However, equation (5) cannot be used directly to inverse eigenvalues, for the number of sine-square terms equals the dimension of matrix that is exponential with $n$. Fortunately, we find that by exploiting the distribution characteristic of the terms of sine-square, equation (5) can be simplified to be

$$\frac{8}{\lambda_j} = \left[ (\sin\frac{\pi}{6})^m \prod_{k=2}^{n-m} \sin(\frac{| 2^k - 2^{-m} j \bmod 2^{k+1} |}{2^{k+1}} \pi) \right]^2 , \qquad (6)$$

where $n \geq 3$, and $m$ is such an integer that lies in $[0, n\text{-}2]$ and makes $2^{-m}j$ to be an odd number. The numerator 8 represents the lower bound of eigenvalues for $n \geq 2$. Details about the derivation of this equation are described in Appendix A. With this equation, the reciprocal of each eigenvalue can be regarded as the product of ($n$-1) terms of sine-square values. And note that sine values can be prepared as probability amplitudes easily through the controlled $R_y$ rotation in quantum computing. This provides the possibility that the amplitude $1/\lambda_j$ in the solution state as shown in equation (4) can be prepared directly by just implementing probability amplitudes all through. The algorithm can thus avoid the need to do phase estimation, Hamiltonian simulation and arithmetic for solving the linear equations of equation (3), and thus for solving the one-dimensional Poisson equation. Furthermore, the complexity of the algorithm is low and the solution error comes only from the initial discretizing process of the Poisson equation.

## 3. Results

### *3.1. Circuits*

The present algorithm proceeds in the following three steps: first with a change of basis, followed by a group of controlled $R_y$ rotations and finally uncomputation. The overall circuit is shown in figure 1.

More specifically, the state $|b\rangle_n$ of register B can be expressed as $\sum_{j=1}^{2^n-1} \beta_j |u_j\rangle$ using the eigenstates of matrix $A$, and the basis is changed into computational basis by $CB$ module as $CB \cdot |b\rangle_n = \sum_{j=1}^{2^n-1} \beta_j | j\rangle$. Then, the ($n$-1) terms of sine-square values in equation (6) are prepared as probability amplitudes of register E using controlled $R_y$ rotations conditioned on the value of $j$ in register B. And then, the reciprocals of eigenvalues are prepared as amplitudes of ancillary qubit through the controlled NOT operation. Next, change the basis in register B back into eigenstates $|u_j\rangle$ to entangle it with the state of ancillary qubit $1/\lambda_j |1\rangle$. Finally, we measure the state of ancillary qubit. If the result is $|1\rangle$, then the state in register B is $\sum_j \beta_j \lambda_j^{-1} |u_j\rangle$ which encodes the solution of the Poisson equation. Typically, the amplitude amplification technique [28] can be applied before measurement to increase the success probability.

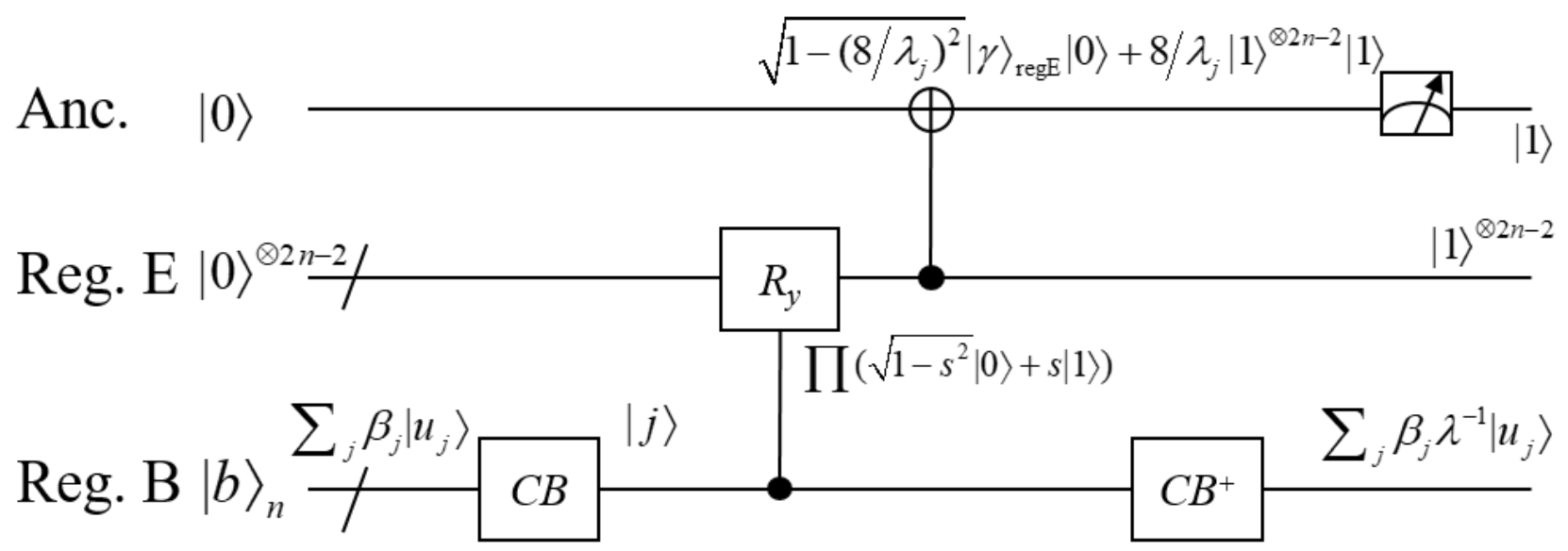

**Figure 1.** The overall circuit of the present QPS. *CB* represents the change of basis from eigenstates $|u_j\rangle$ to computational basis $|j\rangle$. The controlled $R_y$ rotation and NOT operations are used to inverse the eigenvalues on the probability amplitudes, where $s$ denotes the sine values $\sin(j\pi/2^{n+1})$. It is assumed that $|b\rangle_n = \sum_{i=1}^{2^n-1} b_i|i\rangle \Big/ \left\|\sum_{i=1}^{2^n-1} b_i|i\rangle\right\|$ has been prepared and stored in register B.

For the module *CB*, a unitary operator $U$ is required to accomplish the following diagonalization transformation,

$$U|u_j\rangle = |j\rangle. \tag{7}$$

That is, the $j^{\text{th}}$ row of the operator $U$ is $u_j$,

$$U = [u_1, u_2, \ldots, u_{2^n-1}]^T. \tag{8}$$

Apparently, $U$ is the sine transform [16,29], and it can be implemented based on the discrete sine transform (DST) of type I in [29]. The DST is related to discrete Fourier transform (DFT) as follows,

$$T_N^{\dagger} FT_{2N} T_N = C_{N+1}^{I} \oplus (-iS_{N-1}^{I}). \tag{9}$$

The $T_N$ is a basis change matrix, and $FT_{2N}$, $C_{N+1}^{I}$ and $S_{N-1}^{I}$ represents the Fourier, cosine and sine transform, respectively. The circuit to implement the sine transform $S_{N-1}^{I}$ is shown in figure 2.

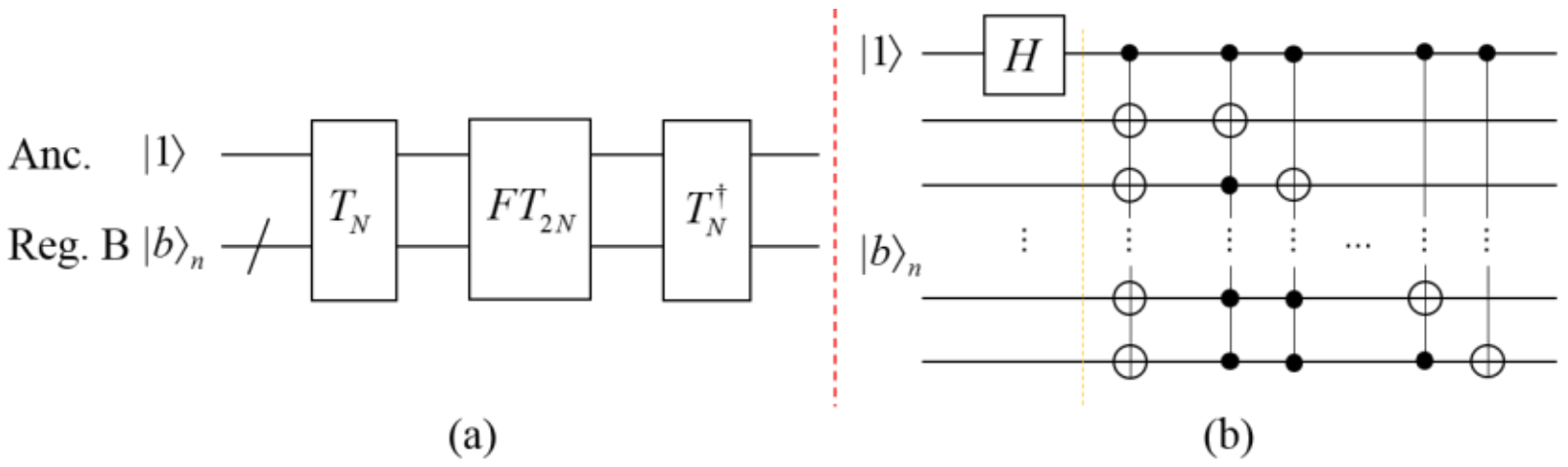

Figure 2. The circuits for the sine transform (a), and for the transform $T_N$ (b). An ancillary qubit set $|1\rangle$ is used to pick out sine transform from the direct sum in equation (9), ignoring the global phase. The circuit costs $n$ qubits and $O(n^2)$ elementary gates after decomposition [30]. Details about this circuit can be found in [18].

Now we turn to the group of controlled $R_y$ rotations to transduce the reciprocals of eigenvalues into a probability amplitude. The reciprocals of eigenvalues are calculated using equation (6). As can be seen, the sine-square terms are indexed by two variables, namely $m$ and $k$. So the overall circuit of this part contains two levels. For the first level, the circuit consists of $n$-1 modules corresponding to $m$ being from 0 to $n$-2. Each module is denoted as $RY_m$ as shown in figure 3. Note that in each module $2^{-m}j$ is an odd number. For the second level, each $RY_m$ module actually implements equation (6) with a certain $m$, thus it contains the ($n$-1) terms of sine-square values. figure 4 takes the $RY_0$ and $RY_1$ modules as the example to show the circuit design of $RY_m$ module. Since sine values can be prepared easily using a single-qubit $R_y$ rotation, the sine-square values are obtained using two qubits as follows,

$$R_y^{\otimes 2}(2\theta)|0\rangle^{\otimes 2} = \cos^2\theta|00\rangle + \cos\theta\sin\theta|01\rangle + \sin\theta\cos\theta|10\rangle + \sin^2\theta|11\rangle. \quad (10)$$

Furthermore, it is very simple to implement the controlled $R_y^{\otimes 2}$ operation in each $RY_m$ module. For a state $|j\rangle$ in register B, the binary representation can be written as $j = j_1 j_2 \cdots j_n = \sum_{k=1}^{n} 2^{n-k} j_k$. Then the $R_y$ rotation can be expressed as

$$R_y(\frac{j\pi}{2^n}) = e^{-i\frac{j\pi}{2^{n+1}}Y} = e^{-i\frac{\pi}{2^{n+1}}(\sum_{k=1}^{n} 2^{n-k} j_k)Y} = \prod_{k=1}^{n} e^{(-i\frac{2^{n-k}\pi}{2^{n+1}}Y)\cdot j_k} = \prod_{k=1}^{n} R_y^{j_k}(\frac{\pi}{2^k}), \quad (11)$$

where $j_k$ is the local control qubit. figure 5 shows the circuit designed for $R_y^{\otimes 2}$ operation based on this equation.

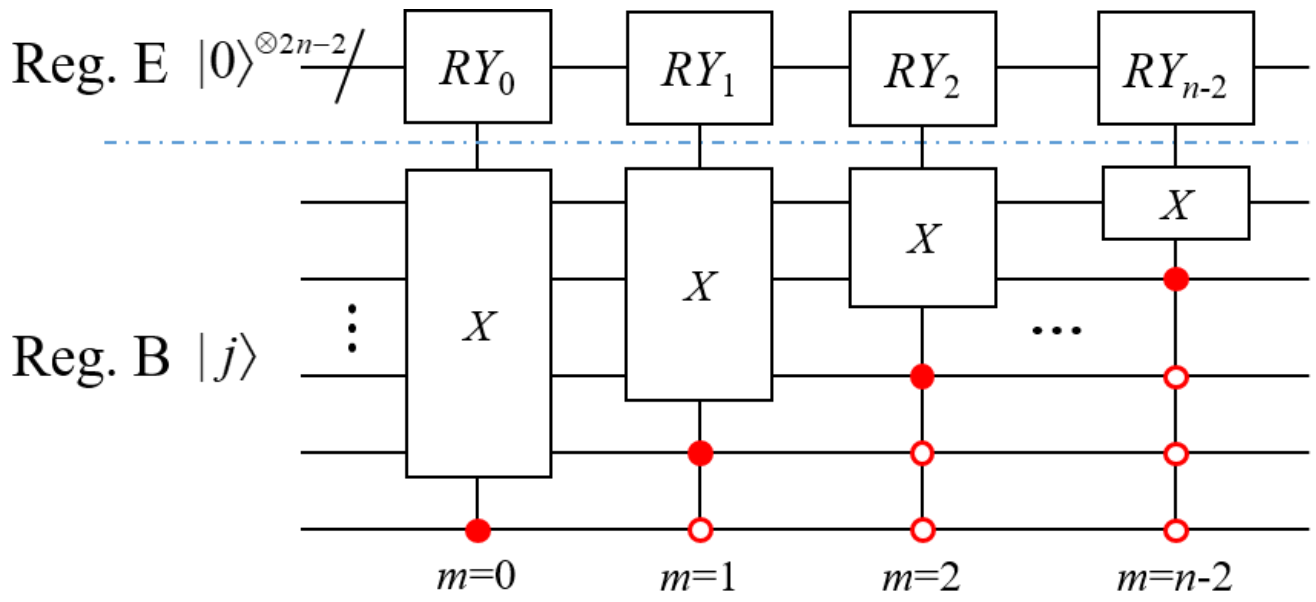

**Figure 3.** The overall circuit to calculate the reciprocal of eigenvalues. In general, the circuit consists of ($n$-1) controlled $RY_m$ modules numbered with $m$ from 0 to $n$-2. Each $RY_m$ module is controlled by two kinds of control qubits called global and local control qubits. The global control qubits are the lower ($m$+1) qubits in Reg. B which are represented by the red dots and circles. The global control qubits are actually used to pick out such $j$ that $2^{-m}j$ is an odd number. The local ones are the higher ($n$-$m$) qubits of Reg. B inside the $X$ modules which will be described below. Note that the ($n$-1) $RY_m$ modules are computed serially. They can turn to be parallel by adding $n$-2 qubits, and then the depth of the circuit is reduced from $O(n^3)$ to $O(n^2)$. A feasible circuit design for such parallelization can be seen in Appendix B.

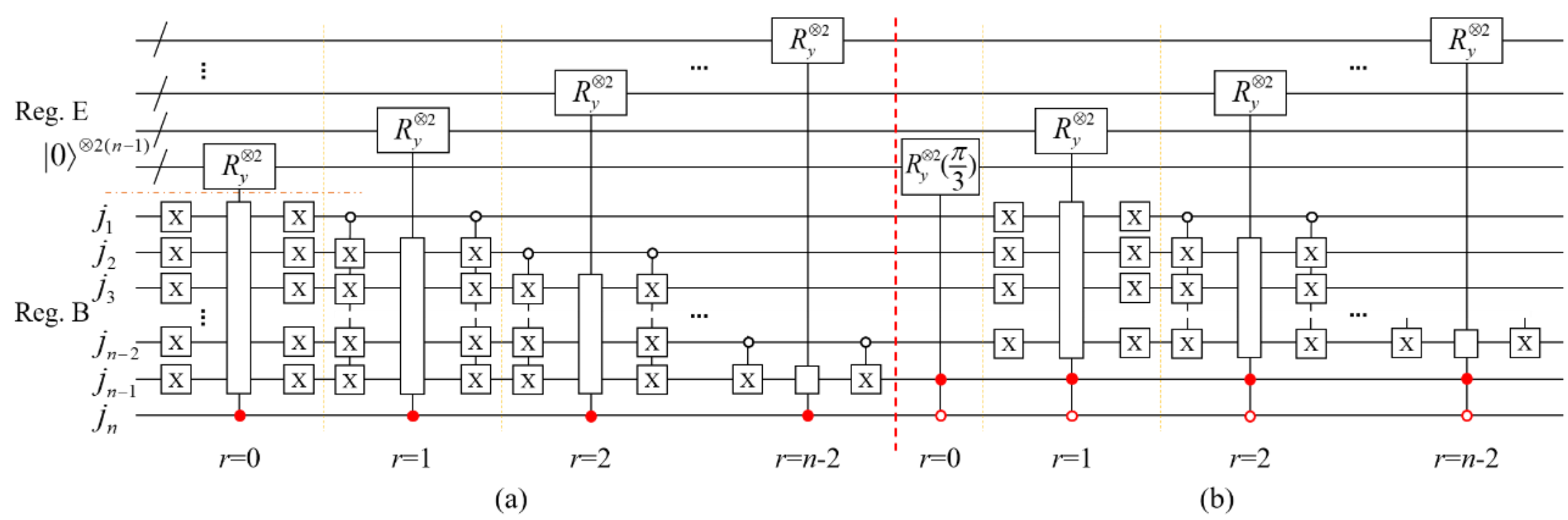

**Figure 4.** The circuit design for $RY_0$ module (a), and $RY_1$ module (b). Each $RY$ module consists of $n$-1 terms of sine-square values. Each $R_y^{\otimes 2}$ operator produce one term of sine-square. The $R_y^{\otimes 2}$ modules are indexed by $r$. For $r$ from 0 to $m$-1, it corresponds to the constant terms of $\sin^2(\pi/6)$; and for $r$ from $m$ to $n$-2, the terms that $k$ being from $n$-$m$ down to 2. The NOT gates are used to calculate the complements of $j$.

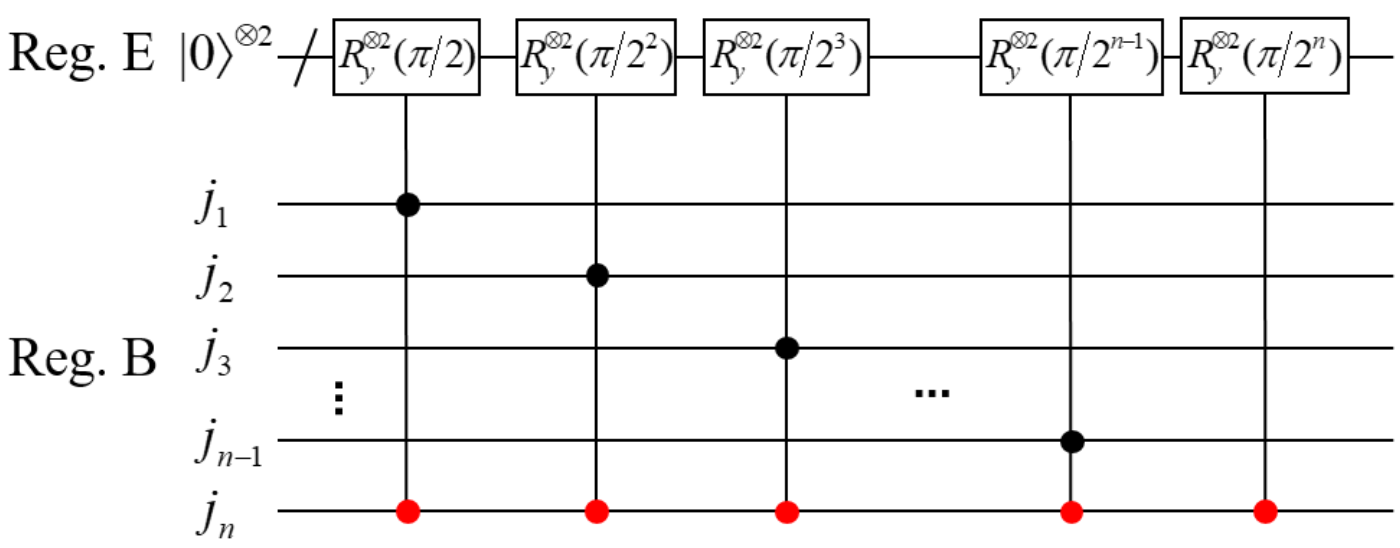

**Figure 5.** The circuit for the first $R_y^{\otimes 2}$ operation in $RY_0$ module in figure 4. The black dots are the local control qubits, and $j_n$ is both global and local control qubit (see the caption of figure 3). According to Lemma 6.1 in [30], one $R_y^{\otimes 2}$ controlled by 2 qubits can be decomposed into 8 two-qubit gates.

After implementing the group of controlled $R_y$ rotations, the state evolves into

$$|0\rangle^{\otimes 2n-2}\sum_{j=1}^{2^n-1}\beta_j|j\rangle \mapsto \sum_{j=1}^{2^n-1}\beta_j\left[(\cos\frac{\pi}{6}|0\rangle+\sin\frac{\pi}{6}|1\rangle)^m\prod_{k=2}^{n-m}(\sqrt{1-s^2}|0\rangle+s|1\rangle)\right]^{\otimes 2}|j\rangle, \quad (12)$$

where $s=\sin(\frac{|2^k-2^{-m}j \bmod 2^{k+1}|}{2^{k+1}}\pi)$. Combining this equation and equation (6), it can be seen that the reciprocals of eigenvalues $8/\lambda_j$ are encoded in the probability amplitudes of the state $|1\rangle^{\otimes 2n-2}$ in register E. Then a multi-controlled NOT operation as shown in figure 1 is used to pick out the state $|1\rangle^{\otimes 2n-2}$ by adding an ancillary qubit. After this operation, the state evolves into

$$\sum_{j=1}^{2^n-1}\beta_j\left[\sqrt{1-(\frac{8}{\lambda_j})^2}|\gamma\rangle_{\text{regE}}|0\rangle+\frac{8}{\lambda_j}|1\rangle^{\otimes 2n-2}|1\rangle\right]|j\rangle, \tag{13}$$

where $|\gamma\rangle$ represents the states of register E except $|1\rangle^{\otimes 2n-2}$.

Finally, the basis is changed back into eigenstates, and the system state evolves into

$$\sum_{j=1}^{2^n-1}\beta_j\left[\sqrt{1-(\frac{8}{\lambda_j})^2}|\gamma\rangle_{\text{regE}}|0\rangle+\frac{8}{\lambda_j}|1\rangle^{\otimes 2n-2}|1\rangle\right]|u_j\rangle. \tag{14}$$

Now, in NISQ devices, we repeatedly measure the ancillary qubit until obtaining state $|1\rangle$ to collapse the system state into $|1\rangle^{\otimes 2n-2}|1\rangle\sum_j\beta_j\lambda_j^{-1}|u_j\rangle$. Alternatively, in the future fault-tolerant quantum devices, before measurement, amplitude amplification can be applied to increase the success probability of obtaining the expected state. That is, when the number of measurement reduced to nearly one, the algorithm become a deterministic one.

### *3.2. Complexity, depth and error*

Now we analyze the complexity, depth and error of our algorithm. As can be seen from figure 1, the algorithm needs $3n$ qubits, where $n$ is the logarithmic of the dimension of the linear system of equations. Considering the ancillary qubits, the number of qubits required is at least $3n+1$. More ancillary qubits can be, of course, added to reduce the complexity of decomposition of gates [30].

The total number of gates required is $5/3n^3$ in one- and two-qubit gates. Furthermore, if the new technology of multi-controlled one-qubit and one-controlled multi-qubit gates can be applied, like the *i*-Toffoli/CNOT$^n$ operations [31], the number of basic gates required can be reduced dramatically.

The depth of the quantum circuit mainly depends on the way of implementing the group of $R_y$ rotations as shown in figure 1. If the $RY_m$ modules are executed serially as shown in figure 3, the depth of the circuit is $5/3n^3$. But the depth can be reduced to $10n^2$ by parallelizing the $RY_m$ modules. The parallelization is achieved by simply adding another $n$-2 qubits. Details are discussed in Appendix B.

Compared to the previous arithmetic-based quantum Poisson solver whose qubit complexity is $22n^2$ and gate $20000n^3$ (only count the highest order) [18], the number of qubits is reduced by one order, and constant factor of the gate complexity is reduced by ten thousand times. This reduction is critical to make the algorithm executable on NISQ computers, and furthermore it reduces dramatically the consumption of the expensive non-Clifford gate on the future fault-tolerant quantum computers with error correction.

Since our algorithm is mainly based on $R_y$ rotations, the major operations are performed on probability amplitudes, which in theory do not cause errors, i.e. without the truncation errors in the phase estimation and arithmetic algorithms. The solution error comes only from the initial finite-difference approximation of the Poisson equation. Recalling that the truncation error is $\varepsilon=1/N^2$, and $n = \log N$, we have $n = 1/2\times\log(1/\varepsilon)$. Therefore, the complexity of our algorithm is $\log(1/\varepsilon)$ in qubits and $\mathrm{poly}(\log(1/\varepsilon))$ in quantum operations.

Since the cost of our circuit is low and the depth can be shallow, the present QPS has the potential to be applied on NISQ devices for solving a non-trivial problem. For example, if solving the one-dimensional Poisson equation by discretizing it with $2^{15}$ points, our QPS can be implemented using 46 qubits and tens of thousands of elementary gates. After parallelization, the depth of the circuit is several thousand with a need of 60 qubits. This cost is acceptable for the NISQ devices with fidelity of 99.99% [33]. The output state encodes the solutions of more than 32 thousand points into the probability amplitudes.

*3.3. Demonstration and execution*

We demonstrate our algorithm using a quantum virtual computing system installed on the Sunway TaihuLight Supercomputer in Wuxi, China [34]. The detailed circuits for the cases of $n$=2 and 3 are shown in figures 6 and 8, respectively.

The case of $n$=2 does not satisfy equation (6) exactly, so the circuit cannot be directly designed based on equation (6). In fact, the circuit in figure 6 is designed based on the very basic idea of our QPS. Specifically, in the case of $n$=2 the three computational basis of 01, 10 and 11 after *CB* module in figure 1 correspond respectively to the eigenvectors of $u_1$, $u_2$ and $u_3$, and they should control the corresponding $R_y$ gates to calculate $8/\lambda_1$, $8/\lambda_2$ and $8/\lambda_3$, respectively. The reciprocals are $8/\lambda_1 = 2\sin^2(2\pi/8)\sin^2(3\pi/8) = \sin^2(3\pi/8)$, $8/\lambda_2 = 2\sin^2(\pi/8)\sin^2(3\pi/8) = \sin^2(\pi/6)$ and $8/\lambda_3 = 2\sin^2(\pi/8)\sin^2(2\pi/8) = \sin^2(\pi/8)$. Furthermore, in figure 6, the double controlled $R_y$ gates are reduced to the single controlled one using the observations: (1) $8/\lambda_2$ would be calculated if and only if the second bit of the basis is 0, regardless of the other bit; (2) $8/\lambda_1$ and $8/\lambda_3$ would be calculated if the second bit is 1. Thus, the calculation of $8/\lambda_1$ is decomposed into two sets, $R_y(\pi/2)$ and $R_y(\pi/4)$, where the $R_y(\pi/4)$ is also used to compute $8/\lambda_3$. That is, $R_y(\pi/2)$ and $R_y(\pi/4)$ are performed when the basis is 01, and $R_y(\pi/4)$ when 11.

The circuit costs only 6 qubits and 70 one- and two-qubit gates without counting the initialization of $b$. When demonstrating the circuit on the virtual system, the state of $|b\rangle_2$ is initialized as $1/\sqrt{2}|01\rangle+1/2|10\rangle+1/2|11\rangle$, and the result is $[0.552987, 0.674065, 0.489736]^T$ after normalized. The expected result calculated using python is $[0.552988, 0.674065, 0.489736]^T$, so the difference is less than $2^{-21}$.

The circuit of figure 6 is optimized further and executed successfully on the near-term small-scale superconducting quantum computers through the cloud of IBMQ [35]; for more details see Appendix C. figure 7 shows the comparison between the experimental and theoretical results. The successful execution makes our QPS the first rudiment of practical differential equation solver for NISQ computers.

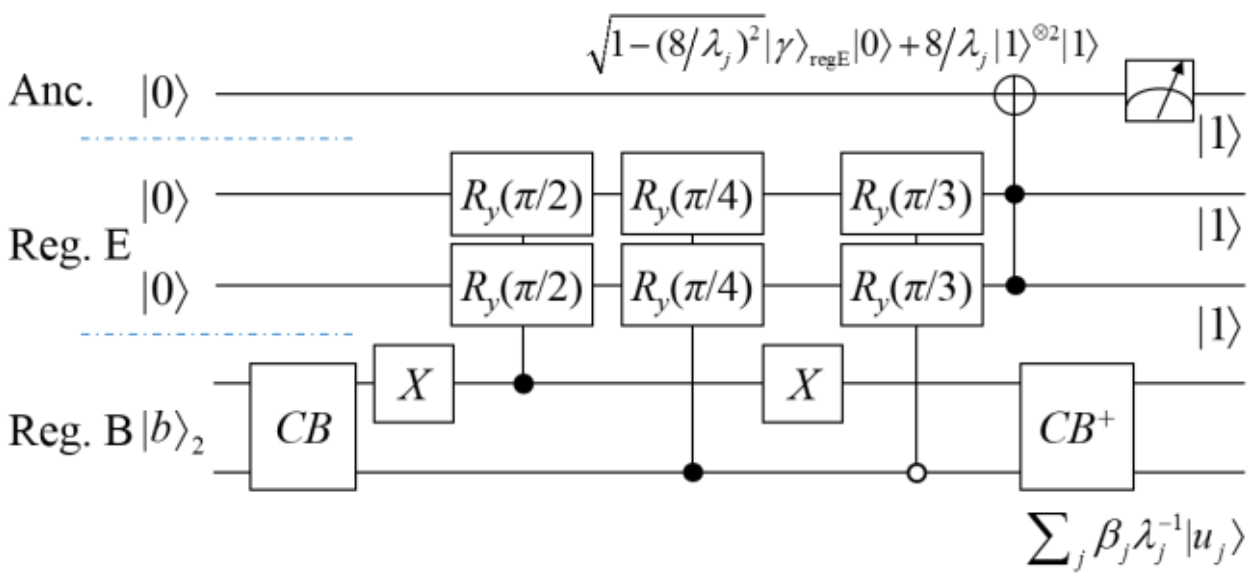

**Figure 6.** The circuit for the case of $n$=2. It needs only 6 qubits and 70 one- and two-qubit gates.

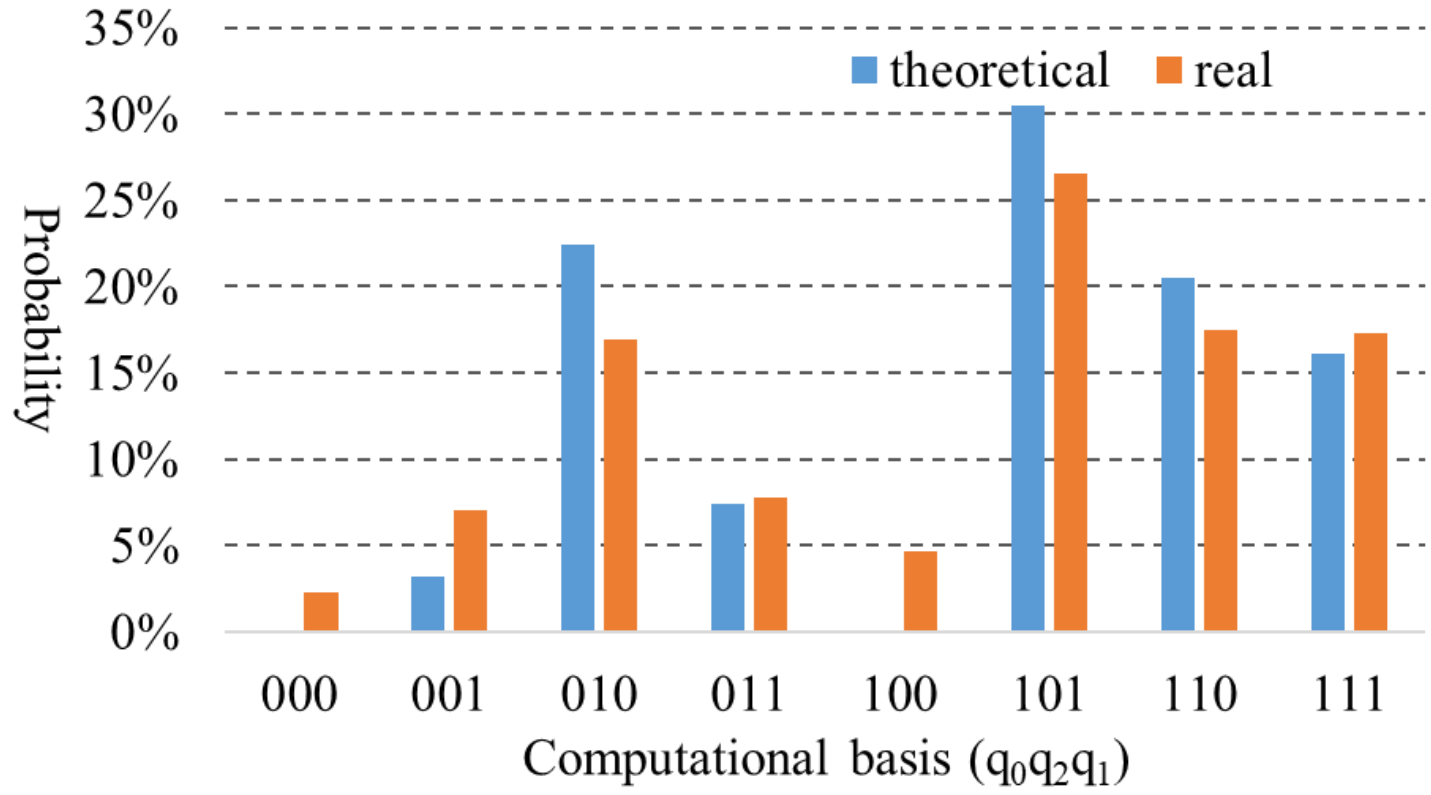

**Figure 7.** The probability distribution of the final states of theory (blue, left bar) and experiment (orange, right bar). The experimental results were obtained by the IBM quantum computer of ibmq_santiago, 27 depth, 4096 shots, calibrated on Nov 29, 2020. The $q_0$ is the ancillary qubit in figure 6. More details about the experiment are provided in Appendix C (the codes are provided in the supplementary material). The error of the probability of the target state is lower than 15%.

Starting from $n$=3, the circuit is designed based on equation (6) in a very direct way, and it scales in a modular way with respect to $n$. We demonstrate the circuits for $n$=3 up to $n$=10 on the virtual system. For the cases of $n\geq 8$, the amplitudes of states corresponding to large eigenvalues cannot be calculated accurately due to the limitation of the virtual system's precision.

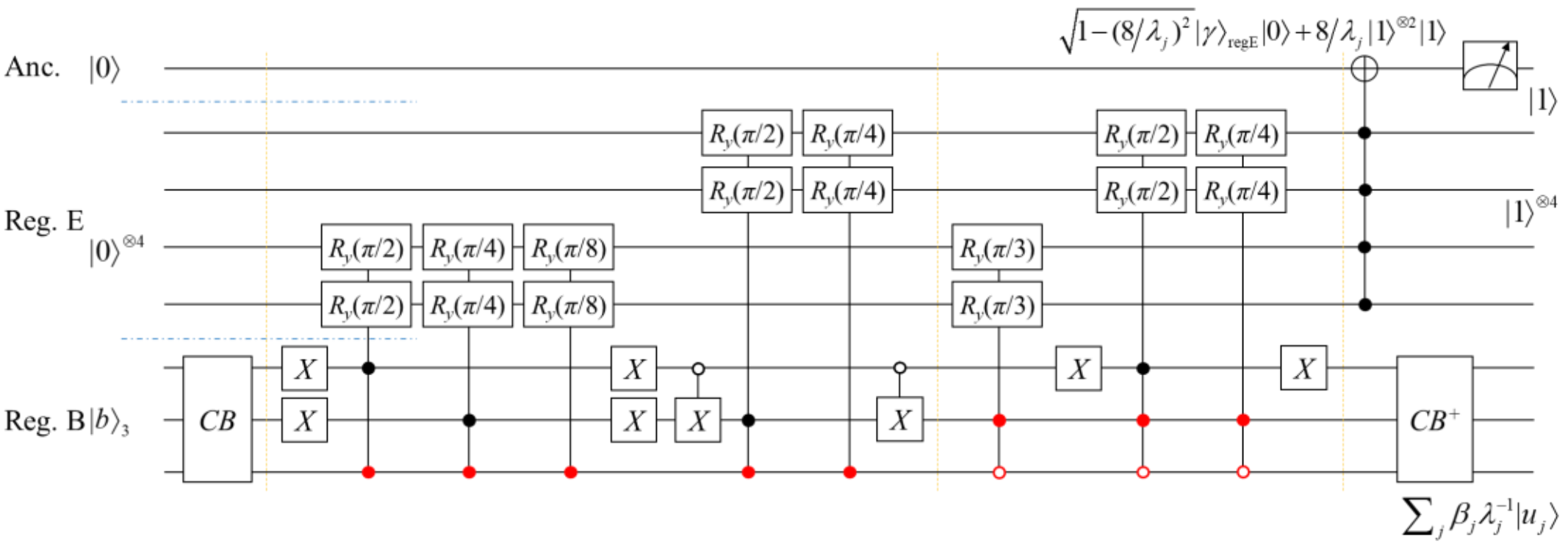

**Figure 8.** The circuit for the case of $n$=3. Being different with that of $n$=2, this circuit is designed in a direct way according to equation (6). Totally, 12 qubits and 200 one- and two- qubit gates are used.

## 4. Conclurions and outlook

We present a compact quantum Poisson solver (QPS) for solving the one-dimensional

Poisson equation. The major operations in the QPS are performed on the probability amplitudes through single-qubit $R_y$ rotation. This is the essential advance of the present QPS compared to the ones based on expensive quantum arithmetic, as well as the phase estimation and Hamiltonian simulation routines. The error of the solution state comes only from the finite difference approximation of the Poisson equation. The idea of implementing function computation on probability amplitude by single-qubit rotation should be inspirational in many ways for optimizing other quantum algorithms.

The costs of our QPS are nearly optimal, that is, $3n$ in qubits and $5/3n^3$ in one- and two-qubit gates, where $n$ is the logarithmic of the number of discrete points. Moreover, the dependence of the complexity on solution error is $\log(1/\varepsilon)$ in qubits and $\mathrm{poly}(\log(1/\varepsilon))$ in quantum operations. When solving the Poisson equation by discretizing it into $2^{15}$ points, our algorithm needs only 46 qubits and tens of thousands of one- and two-qubit gates, and the error of the solution state is, in theory, $2^{-30}$. So the present QPS would be capable of running on near-future Noisy Intermediate-Scale Quantum (NISQ) devices for real applications.

An open question is how to extend the present algorithm to $d$-dimensional Poisson equations. For higher dimensions, the eigenvalues are the sum of that of one-dimensional case. We think there should be an intuitive way to inverse the eigenvalues on probability amplitudes, like the technique of quantum signal processing [36], and achieve a complexity with nearly linear dependence on dimensions.

**Acknowledgements**

We are very grateful to the National Supercomputing Center in Wuxi for the great computing resource. We would also like to thank the technical team from the Origin Quantum Computing Technology co., LTD in Hefei for the professional services on quantum virtual computation. We acknowledge the use of IBM Quantum services for this work. The views expressed are those of the authors, and do not reflect the official policy or position of IBM or the IBM Quantum team. The present work is financially supported by the National Natural Science Foundation of China (Grant No. 12005212, 61575180) and the Pilot National Laboratory for Marine Science and Technology (Qingdao).

**Data availability statement**

All data that support the findings of this study are included within the article (and any supplementary files).

**Appendix A. Reciprocal of eigenvalues**

We use central-difference-approximation to discretize the one-dimensional Poisson equation, and the resulted linear system of equations is solved by a special QLSA. The discretized matrix $A$ is a Cartan matrix, and the product of all its eigenvalues satisfies the so-called sine formula as shown in equation (5) [22]. This equation shows that the reciprocal of each eigenvalue equals to the product of the remained $2^n$-2 eigenvalues. The number of terms can be reduced exponentially as shown in equation (6) in the following way.

Firstly, we study the distribution characteristics of the terms on the left hand side of the following equation derived from equation (5),

$$\prod_{j=1}^{2^n-1} \sin^2\left(\frac{j}{2^{n+1}}\pi\right) = 2^{n+2-2^{n+1}}. \tag{A.1}$$

The angular coefficient $j/2^{n+1}$ of each term is listed in the way as shown in figure A1. Apparently, each layer contains $2^{n-1}$ odd terms that $j$ is an odd number, and $2^{n-1}$-1 even terms that $j$ is an even number. All of the even terms in one layer are actually the ones of the adjacent upper layer, and each even term always corresponds to an odd one of some upper layer. That is, all the terms of one layer, say $n$=4 layer, are actually the odd terms of the present and all upper layers, i.e. layers of $n$=4, 3, 2, 1. So all the terms are divided into $n$ groups and each group contains only the odd terms of that layer.

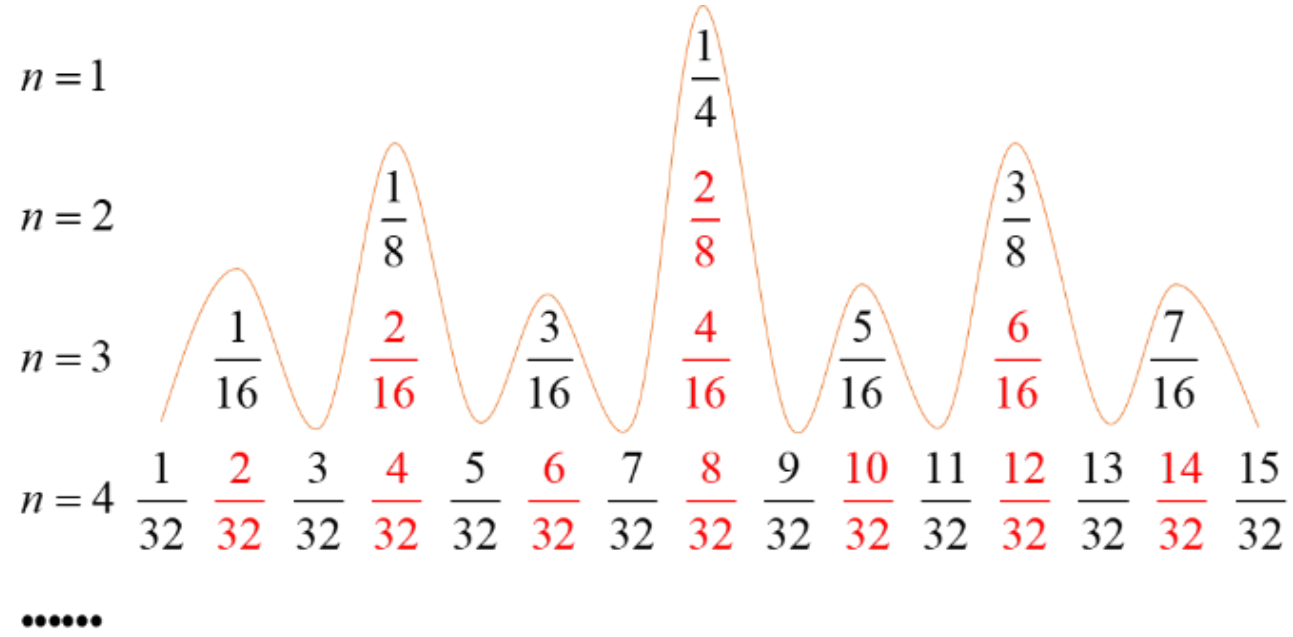

**Figure A1.** The distribution of the angular coefficient $j/2^{n+1}$ of each term in equation (5) for $n$=1, 2, 3, 4. The characteristic of distribution is that the terms of one layer consist of all the odd terms of the present and all upper layers.

Using the above distribution characteristic, the product can be re-organized with only the odd terms of each layer as follows,

$$\prod_{k=1}^{n} \prod_{j=1}^{2^{k-1}} \sin^2 \frac{(2j-1)\pi}{2^{k+1}} = 2^{n+2-2^{n+1}}. \tag{A.2}$$

Secondly, we analyze the reciprocal of each term $\sin^2(j\pi/2^{n+1})$. The following discussions are for two conditions, one for $j$ being an odd number and the other for $j$ being an even number.

When $j$ is an odd number belonging to [1, $2^n$-1], starting from equation (A.1) we have

$$
\begin{aligned}
\sin^{-2}\left(\frac{j}{2^{n+1}}\pi\right) &= 2^{2^{n+1}-n-2}\prod_{\substack{k=1\\ k\neq j}}^{2^n-1}\sin^2\left(\frac{k}{2^{n+1}}\pi\right) \\
&= 2^{2^{n+1}-n-2}\sin^2\left(\frac{2^n-j}{2^{n+1}}\pi\right)\prod_{\substack{k=1\\ k\neq j\\ k\neq 2^n-j}}^{2^n-1}\sin^2\left(\frac{k}{2^{n+1}}\pi\right) \\
&= 2^{2^{n+1}-n-2}\sin^2\left(\frac{2^n-j}{2^{n+1}}\pi\right)\sin^2\left(\frac{2^{n-1}}{2^{n+1}}\pi\right)\cdot\prod_{\substack{k=1\\ k\neq j\\ k\neq 2^n-j}}^{2^{n-1}-1}\left[\sin^2\left(\frac{k}{2^{n+1}}\pi\right)\sin^2\left(\frac{2^n-k}{2^{n+1}}\pi\right)\right] \\
&= 2^{2^{n+1}-n-2}\sin^2\left(\frac{2^n-j}{2^{n+1}}\pi\right)\sin^2\left(\frac{2^{n-1}}{2^{n+1}}\pi\right)\cdot\prod_{\substack{k=1\\ k\neq j\\ k\neq 2^n-j}}^{2^{n-1}-1}\left[\sin^2\left(\frac{k}{2^{n+1}}\pi\right)\cos^2\left(\frac{k}{2^{n+1}}\pi\right)\right] \\
&= 2^{2^{n+1}-n-2}\sin^2\left(\frac{2^n-j}{2^{n+1}}\pi\right)\sin^2\left(\frac{2^{n-1}}{2^{n+1}}\pi\right)\cdot 2^{4-2^n}\prod_{\substack{k=1\\ k\neq j\\ k\neq 2^n-j}}^{2^{n-1}-1}\sin^2\left(\frac{k}{2^n}\pi\right) \\
&= \cdots\cdots \\
&= 2^{2^{n+1}-n-2}\left(\prod_{k=2}^{n}2^{4-2^k}\right)\cdot\left[\prod_{k=2}^{n}\sin^2\left(\frac{2^{k-1}}{2^{k+1}}\pi\right)\right]\cdot\left[\prod_{k=2}^{n}\sin^2\left(\frac{|2^k-j \bmod 2^{k+1}|}{2^{k+1}}\pi\right)\right] \\
&= 2^{2n-1}\cdot\prod_{k=2}^{n}\sin^2\left(\frac{|2^k-j \bmod 2^{k+1}|}{2^{k+1}}\pi\right)
\end{aligned}
\quad . \text{(A.3)}
$$

The above derivation contains a recursive process that cut the number of terms of the product by half utilizing the symmetry of distribution in each layer as shown in figure A1. The two trigonometric formulas, i.e. $\sin 2\theta = 2\sin\theta\cos\theta$ and $\sin\theta = \cos(\pi/2-\theta)$, are used repeatedly in the derivation. So the reciprocal of eigenvalues with $j$ being an odd number can be calculated by the ($n$-1) terms as follows,

$$
\frac{1}{\lambda_j} = \frac{1}{2^{2n+2}\sin^2(j\pi/2^{n+1})} = 2^{-3}\prod_{k=2}^{n}\sin^2\left(\frac{|2^k-j \bmod 2^{k+1}|}{2^{k+1}}\pi\right). \quad \text{(A.4)}
$$

When $j$ is an even number, it can be transformed to an odd one in some upper layer. Suppose $j = 2^m i$, where $i$ is an odd number and $m$ is an integer belonging to [1, $n$-2], then the even term $j$ corresponds to the odd term $i$ in the upper $m^{\text{th}}$ layer. In such case, $\sin^{-2}(j\pi/2^{n+1})$ can be calculated as

$$
\sin^{-2}\left(\frac{j}{2^{n+1}}\pi\right) = \sin^{-2}\left(\frac{i}{2^{n-m+1}}\pi\right) = 2^{2(n-m)-1}\cdot\prod_{k=2}^{n-m}\sin^2\left(\frac{|2^k-2^{-m}j \bmod 2^{k+1}|}{2^{k+1}}\pi\right). \quad \text{(A.5)}
$$

In fact, equation (A.3) for odd $j$ can be seen as a particular case of equation (A.5) with $m$ = 0. So, in summary, the reciprocal of eigenvalues for any $j$ can be calculated by

$$
\frac{1}{\lambda_j} = \frac{1}{2^{2n+2}\sin^2(j\pi/2^{n+1})} = 2^{-3}\cdot 4^{-m}\prod_{k=2}^{n-m}\sin^2\left(\frac{|2^k-2^{-m}j \bmod 2^{k+1}|}{2^{k+1}}\pi\right). \quad \text{(A.6)}
$$

The lower bound of eigenvalues equals to 8 when $n \geq 2$. So after transposing $2^{-3}$ to the left side of equation (A.6) and transforming the factor $4^{-m}$ into a sine form, equation (A.6) turns to

$$\frac{8}{\lambda_j}=\left[(\sin\frac{\pi}{6})^m\cdot\prod_{k=2}^{n-m}\sin(\frac{|2^k-2^{-m}j \bmod 2^{k+1}|}{2^{k+1}}\pi)\right]^2. \tag{A.7}$$

This is pretty intriguing. The reciprocal of $8/\lambda_j$ is the product of ($n$-1) terms and each term is a square of sine value. The first $m$ terms are constants, namely $\sin^2(\pi/6)$. Put it another way, we can inverse the eigenvalues through ($n$-1) sine-square terms. The number of terms of the product is reduced exponentially from ($2^n$-2) to ($n$-1), and this guarantees that the complexity of our algorithm is $3n$ in qubits.

## Appendix B. Parallelization of $RY_m$ modules

In figure 3, the $RY_m$ modules are executed serially. The gate-complexity of the circuit is low, but the depth is high to run efficiently on NISQ devices. Here we propose a feasible way of paralleling the $RY_m$ modules to reduce the depth from $O(n^3)$ to $O(n^2)$. The parallelization consists of three aspects as follows.

Firstly, parallelize the control qubits in figure 3. Another register $C$ with ($n$-2) qubits is allocated to store the global control qubits. The control qubits are parallelized by the *CP* (Control-qubits Parallelization) module as shown in figure B1.

Secondly, break up each $RY_m$ module in figure 4 and assemble new module called *RYP* (*RY* Parallelization) based on the basic unit, the controlled $R_y^{\otimes 2}$ operation. Let $R_y(m,r)$ represent the controlled $R_y^{\otimes 2}$ indexed by $m$ and $r$ as shown in figures 3 and 4, respectively. The assembling rule is that the $R_y(0,r)$, $R_y(1, r\text{+}1\bmod(n\text{-}1))$, …, $R_y(m, r\text{+}m \bmod(n\text{-}1))$, …, and $R_y(n\text{-}2, r\text{+} n\text{-}2 \bmod(n\text{-}1))$ are combined to form the module $RYP_r$. figure B1 shows the circuit for $RYP_0$ and $RYP_1$. In such way, all the $R_y^{\otimes 2}$ are rearranged to be parallel in each *RYP* module.

Thirdly, as shown in figure 5, the controlled $R_y^{\otimes 2}(\pi/2^i)$ in each *RYP* are also rearranged to be parallel. The basic idea is illustrated in figure B2 for $RYP_0$ module.

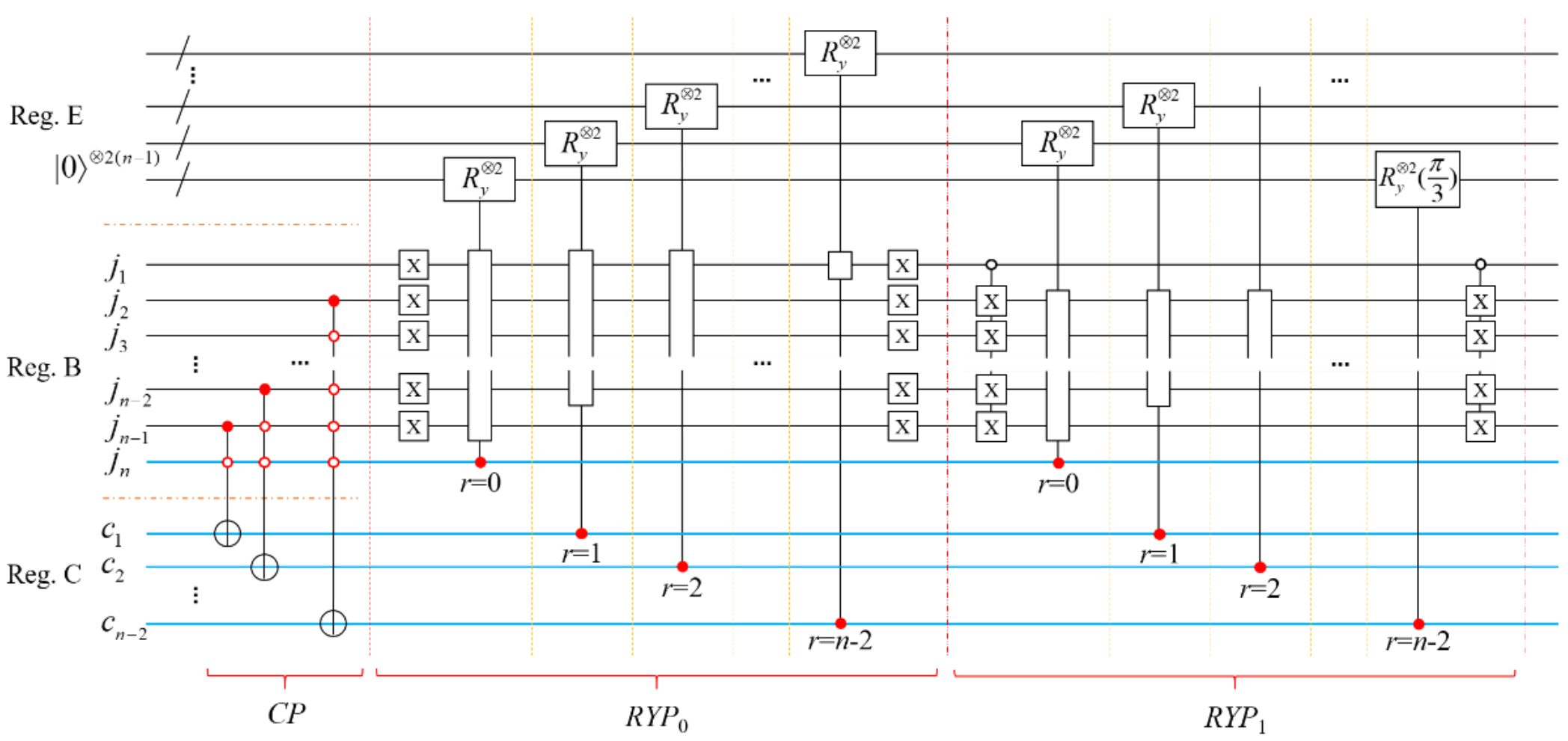

**Figure B1.** A feasible way of parallelizing the *RY* modules in figure 3. The *CP* module is used to parallelize the control qubits and *RYP* represent the parallelized *RY* modules. The depth of *CP* is $5(n\text{-}2)^2$ when the multi-controlled NOT gates are decomposed into one- and two-qubit gates.

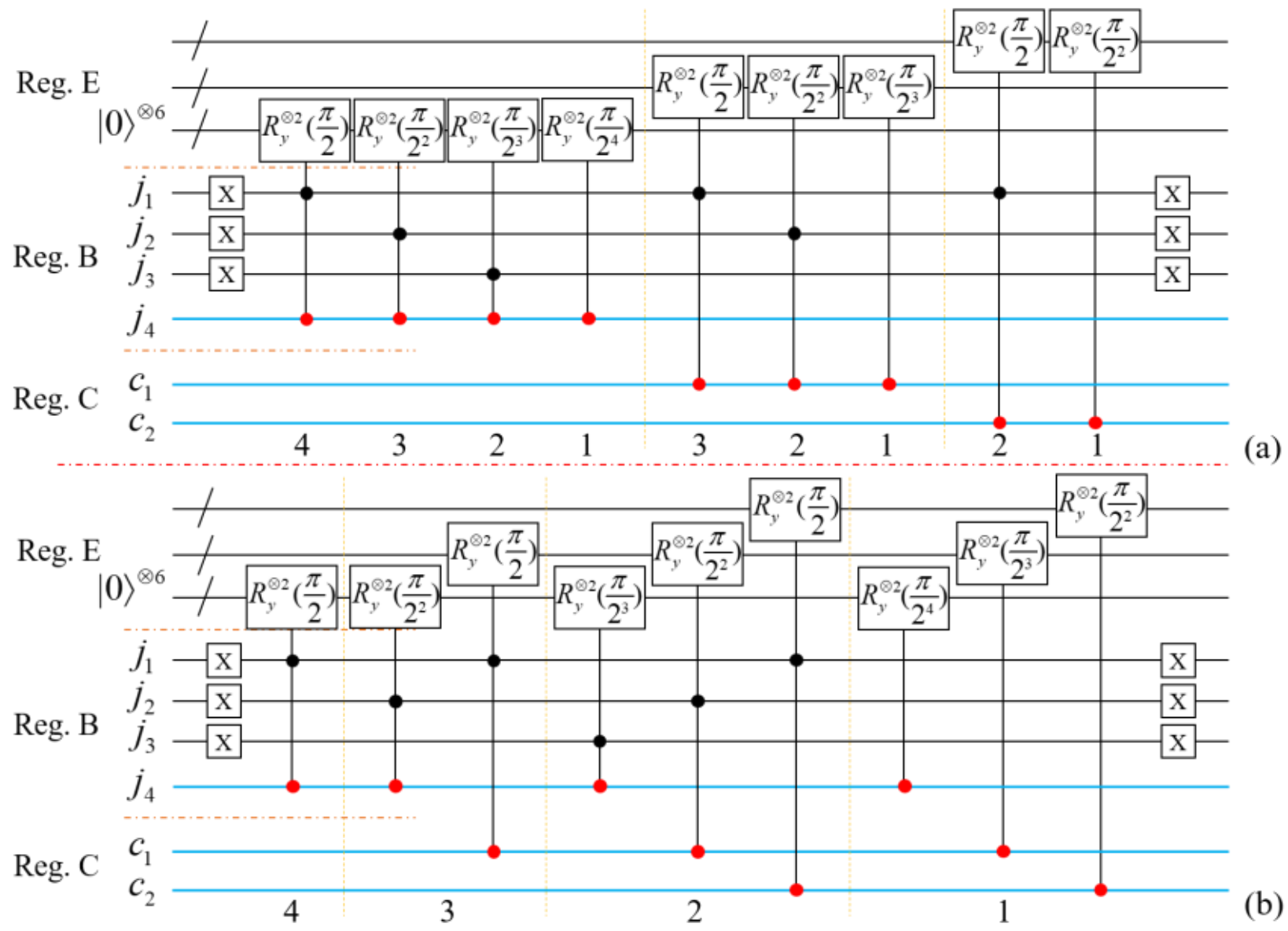

**Figure B2.** The way of parallelizing the controlled $R_y^{\otimes 2}(\pi/2^i)$ in $RYP_0$ module with $n$=4. (a) The original arrangement, and (b) the re-arrangement. The controlled $R_y^{\otimes 2}(\pi/2^i)$ with the same number labeled below the circuit in (a) is rearranged together in (b). The depth of the $RYP_0$ is reduced from $4n^2$ to $8n$ when decomposed by one- and two-qubit gates. The numbers imply that the depth of *RYP* depends on $RY_0$. The total depth of *RYP* modules is $5n^2$.

As can be seen, the depth of the parallel-version QPS mainly depends on the $RY_0$ and *CP* modules. The gate-complexity of $RY_m$ modules turns to $4n$ in qubits and the depth reduces from $5/3n^3$ to $10n^2$. When $n$=15, the depth of the circuit inversing eigenvalues based on serial and parallel $RY_m$ modules is about 8000 and 1800, respectively. Since the decomposition of the multi-controlled one-qubit and one-controlled multi-qubit gates constitute the major cost, the depth of $T_N$ in *CB* module (in figure 2) and *CP* in $RY_m$ can be reduced to $O(n)$ using the technology of *i*Toffoli/CNOT$^n$ gates [31]. Additionally, the depth of $FT_{2N}$ in *CB* can also be reduced to $O(n)$ even $O(\log n)$ [37,38].

## Appendix C. Execution on IBM quantum computers

Today's gate-based non fault-tolerant quantum computers that has noise and limited resources eclipse the realization of certain quantum algorithms because it can only provide limited width and depth for the reliable execution of the algorithms on it, among which is the famous quantum linear system problem (QLSP) [33,39]. Here, with several deeper optimizations, we show that the basic case of $n$=2 of our QPS can be executed successfully on the near-term superconducting quantum computers publicly provided by IBMQ. The width and depth of the circuit is lowed to 3 and 21, respectively. The optimization mainly comprise the following two aspects: (1) decomposition of the *CB* module, (2) calculation of the eigenvalue reciprocal (the controlled $R_y$ rotations).

### *C.1. Decomposition of the CB module*

The original $CB_{3\times3}$ operator has well scalability, but the complexity is slightly high leading to the poor executability on the near-term quantum computers. Here $CB_{3\times3}=[u_1,u_2,u_3]$ is expanded to $CB_{4\times4}=[u_0,u_1,u_2,u_3]$ that is a special unitary matrix, namely SU(4), through appending $u_0=[-1,0,0,0]^T$ orthogonal to other $u_j$. Higher dimensional $CB$ can also be constructed in a similar way, such as $CB_{8\times8}$ for $n$=3. According to theorems 1 and 3 in [40], we obtain the following operator $U$ in the magic basis, i.e. $U = M \cdot CB_{4\times4} \cdot M^{\dagger}$

$$U = \begin{pmatrix} \frac{1}{2} & -\frac{\sqrt{2}-i}{2} \\ -\frac{\sqrt{2}+i}{2} & -\frac{1}{2} \end{pmatrix} \otimes \begin{pmatrix} -\frac{1}{2} & \frac{\sqrt{2}+i}{2} \\ \frac{\sqrt{2}-i}{2} & \frac{1}{2} \end{pmatrix}. \tag{A.8}$$

Operator $U$ then is implemented by two $U_3$ gates as $U_3(\theta, \phi, \lambda)\otimes U_3(\alpha, \beta, \gamma)$ on IBM quantum computers with $\theta=2\pi/3$, $\phi=1.1959\pi$, $\lambda=-0.1959\pi$ and $\alpha=4\pi/3$, $\beta=-0.1959\pi$, $\gamma=1.1959\pi$, respectively.

Finally, by this means, the transpiled circuit of $CB$ module is optimal with 5 depth and 2 qubits as shown in figure C1.

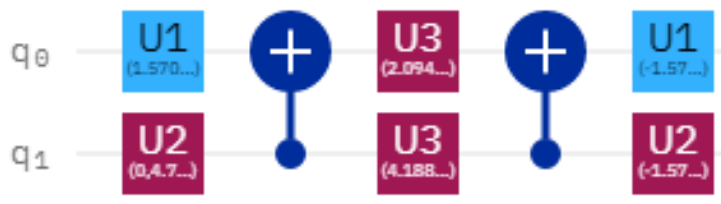

**Figure C1.** The decomposition and execution of the 4×4 $CB$ module on IBM quantum computers.

### *C.2. Calculation of the eigenvalue reciprocals*

The main text shows that the calculation of each $8/\lambda_j$ is carried out by two $R_y$ gates in figure 6, namely two $R_y$ gates produce a term of sine square. This circuit has well scalability, but the complexity is slightly high. To simplify the execution of the circuit, the two controlled $R_y$ gates is shrunk to single one, namely $\sin\theta_j=\sin^2\phi_j=8/\lambda_j$. The optimized circuit is shown in figure C2. Both the number of controlled $R_y$ gates and qubits is reduced to 3.

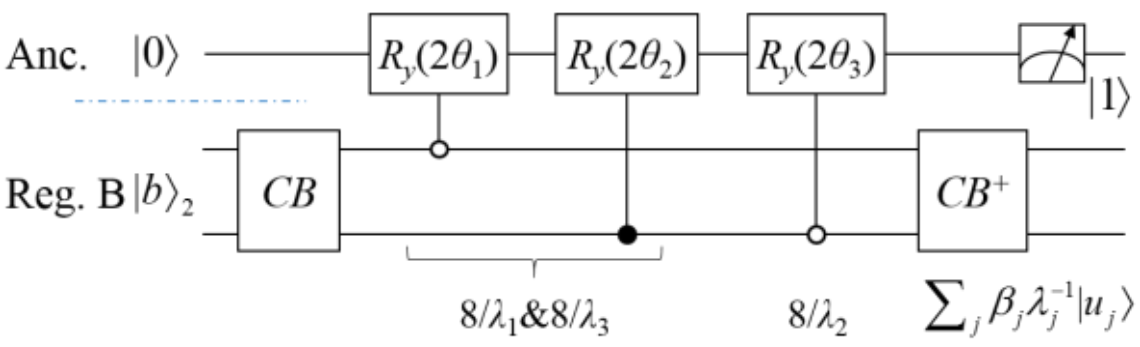

**Figure C2.** The overall circuit of the optimized 3-qubit QPS for $n$=2. The parameter $\theta_j$ satisfies $\sin\theta_2=\sin^2(\pi/8)$, $\sin\theta_3=\sin^2(\pi/6)$ and $\theta_1=\arcsin[\sin^2(3\pi/8)] - \arcsin[\sin^2(\pi/8)]$ as we discussed above. Note that each controlled $R_y$ gates need to be decomposed into 4 native gates to execute.

Up to now, the logical circuit is optimal. However the tranpiled circuit need to be taken care for their diverse architectures, namely the different connectivity types between qubits in different real backends [41,42]. The 2-qubit operations can merely perform on the two qubits with direct connectivity, while the indirectly connected qubits need to move to be adjacent through SWAP gates first. For the circuit of our QPS, after transpiled on different backends, the depth is 23 on ibmq_santiago and ibmq_vigo with limited connectivity and 21 on ibmqx2 with fully connected qubits (qubits 0,1,2 or 2,3,4 as shown in figure C4). The circuits are shown in figure C3.

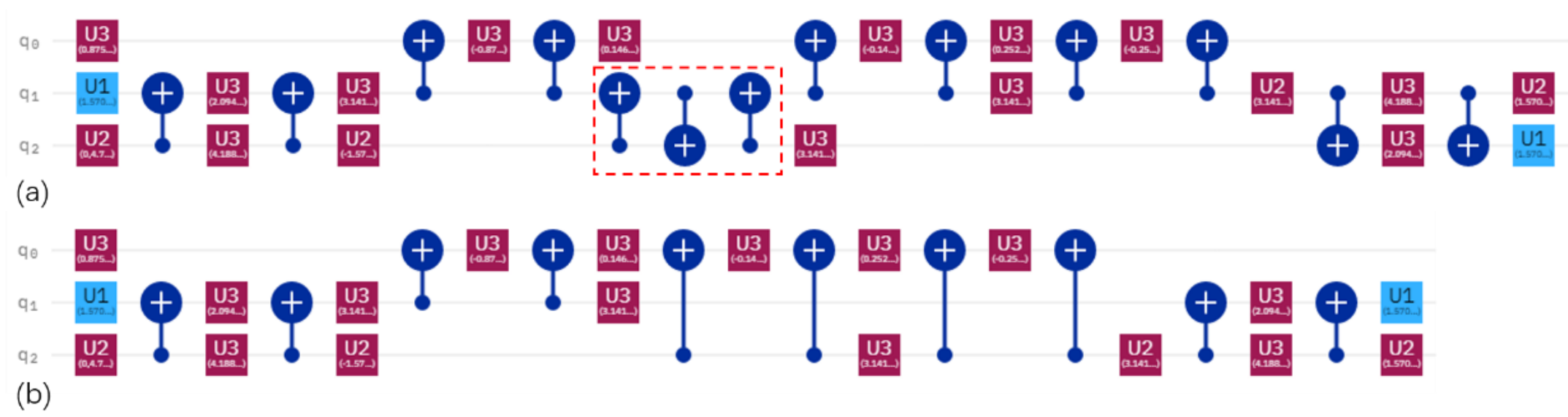

**Figure C3.** The tranpiled circuits on ibmq_santiago (a) and ibmqx2 (b). A SWAP is introduced, as shown in the red grid in (a), to connect the physical qubits of $q_0$ and $q_2$ through the middle $q_1$ while the fully connected one shown in (b) needs no SWAP.

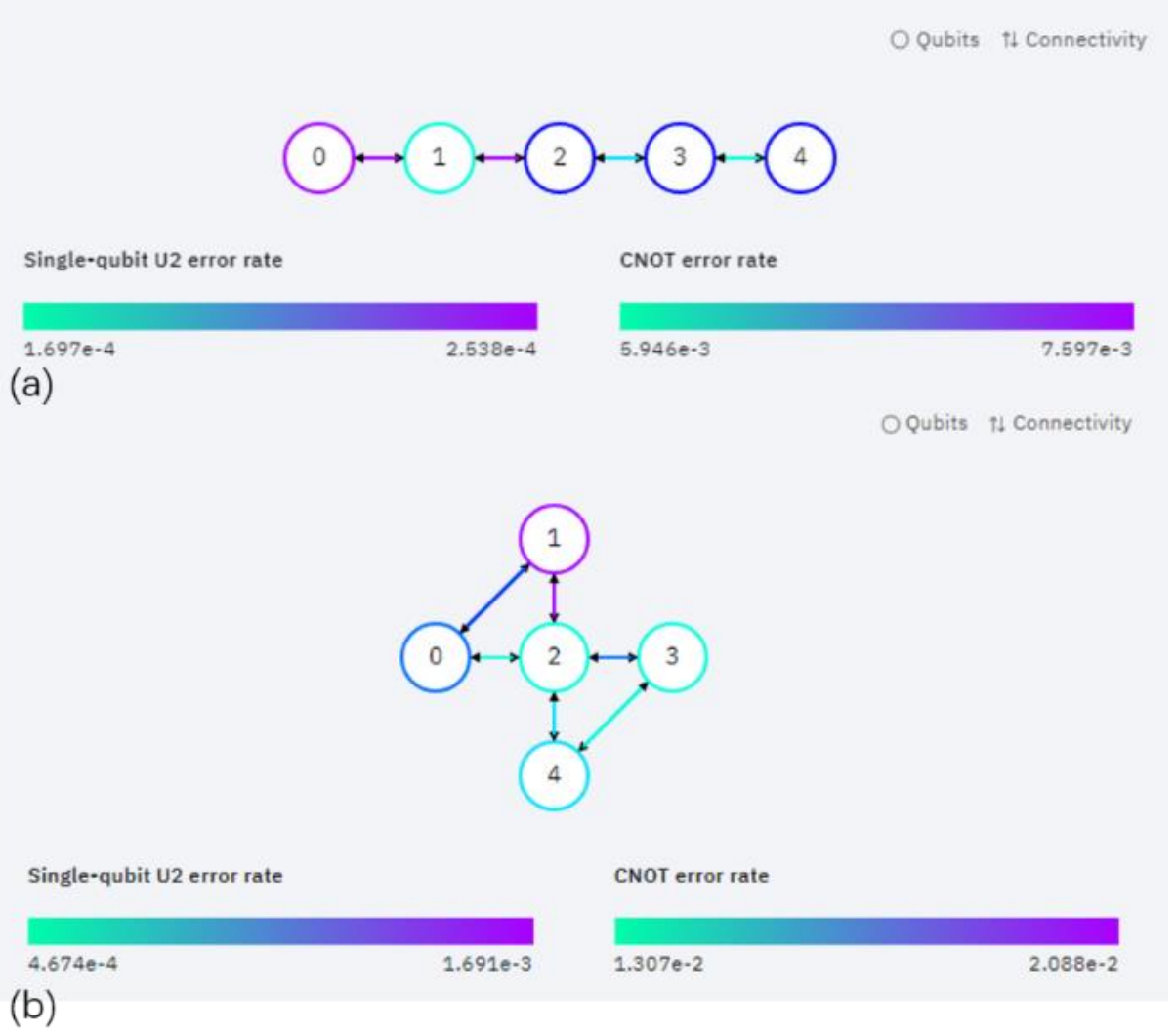

**Figure C4.** The qubit architecture and error rates of ibmq_santiago (a) and ibmqx2 (b). The calibration time is Nov 27 and 24, 2020, respectively.

Note that although the circuit for ibmq_santiago has 3 more CNOT gates than that for ibmqx2, ibmq_santiago is still a better choice for us because of its much smaller error rate relative to ibmqx2 as shown in figure C4.

The CNOT error rate $\varepsilon$ of ibmq_santiago is about 0.006 among 2, 3 and 4 while ibmqx2 is 0.02. In fact, ibmq_santiago is far from reaching the critically executable relation of $w{\cdot}d \ll 1/\varepsilon$, but it is sufficient for us to obtain a solution that is consistent with the theoretical results as shown in figure 7 in the main text.

Furthermore, the error of the probability of the target state is reduced to less than 10%

by the technique of leakage elimination [44] while the depth of the circuit remains unchanged. As aforementioned, the logical circuit is optimal in each module, so the depth of the entire circuit can be reduced to less than 20 based on the methods for two-even three-qubit operator decompositions [40,43,45].

# Supplementary online material for the paper

## *Quantum Poisson Solver without Arithmetic*

Shengbin Wang[1,†], Zhimin Wang[1,†], Guolong Cui[1], Shangshang Shi[1], Ruimin Shang[1], Jiayun Li[1], Wendong Li[1], Zhiqiang Wei[2,3,*] and Yongjian Gu[1,*]

[1] College of Physics and Optoelectronic Engineering, Ocean University of China, Qingdao 266100, China
[2] College of Computer Science and Technology, Ocean University of China, Qingdao 266100, China
[3] High Performance Computing Center, Pilot National Laboratory for Marine Science and Technology (Qingdao), Qingdao 266100, China
[†] These authors contributed equally: Shengbin Wang, Zhimin Wang
[*] Author to whom any correspondence should be addressed.
E-mail: yjgu@ouc.edu.cn; weizhiqiang@ouc.edu.cn

Here we provide the codes for the case of *n*=2 and 3 on the virtual system, and the code for *n*=2 on ibmq_santiago.

### 1. Simulation codes

### The code for *n*=2

```
%the virtual system has its basic gates of H, X, Ry, CR, CNOT, SWAP, TOFFOLI, etc
%they are converted to one & two qubit gates to get the number of 70 (indeed, in this specific case, the total number of gates is 70, including the 3 gates for the initialization of b. So we take the integer 70)

%6 qubits and 70 one & two qubit gates
%qubits 0-2 regB
%qubits 3-4 regE
%qubit 5 register Anc.

%state b preparation
%%3 one & two qubit gates %%
H 1
X 1
CNOT 1,0
X 1

RY 0,"pi/4"
CNOT 1,0
```

```
RY 0,"-pi/4"
%state b preparation done

%base conversion CB (sine transform)
%%26 gates%%
X 2
H 2
CNOT 2,0
CNOT 2,1
TOFFOLI 0,2,1
CNOT 2,0
%%%QFT
H 2
CR 1,2,"pi/2"
CR 0,2,"pi/4"
H 1
CR 0,1,"pi/2"
H 0
SWAP 0,2

DAGGER
H 2
```

```
CNOT 2,0
CNOT 2,1
TOFFOLI 0,2,1
CNOT 2,0
ENDDAGGER

%base conversion CB (sine transform) done
%%%%%%%%%%%%%%%%%%%%%
%%10 gates%
%the calculation Ry
X 1
CNOT 1,3
CNOT 1,4
RY 3,"-pi/4"
RY 4,"-pi/4"
CNOT 1,3
CNOT 1,4
RY 3,"pi/4"
RY 4,"pi/4"

CNOT 0,3
CNOT 0,4
RY 3,"-pi/8"
RY 4,"-pi/8"
CNOT 0,3
CNOT 0,4
RY 3,"pi/8"
RY 4,"pi/8"

X 1

X 0
CNOT 0,3
CNOT 0,4
RY 3,"-pi/6"
RY 4,"-pi/6"
CNOT 0,3
CNOT 0,4
RY 3,"pi/6"
RY 4,"pi/6"
X 0

%the last TOFFOLI gate
TOFFOLI 3,4,5

DAGGER
%base conversion CB (sine transform)
X 2
H 2
CNOT 2,0
CNOT 2,1
TOFFOLI 0,2,1
CNOT 2,0
%%%QFT
H 2
CR 1,2,"pi/2"
CR 0,2,"pi/4"
H 1
CR 0,1,"pi/2"
H 0
SWAP 0,2

DAGGER
H 2
CNOT 2,0
CNOT 2,1
TOFFOLI 0,2,1
CNOT 2,0
ENDDAGGER

%base conversion CB (sine transform) done
ENDDAGGER

PMEASURE 5,1,0
```

## The code for *n*=3

```
%12 qubits and 200 one & two qubit gates
%qubits 0-3 regB
%qubits 4-7 regE
%qubit 8 register Anc.
%qubits 9-11 the ancillarys

%b3
%
X 0
H 0
```

```
H 1
H 2
%
X 1
X 2
RY 0,"pi/4"
TOFFOLI 2,1,0
RY 0,"-pi/4"
X 1
X 2
%%%%%%%%%%%%%%%%
%CB
%
X 3
%%%%%%%%%%%%%%%%
H 3
CNOT 3,2
CNOT 3,1
CNOT 3,0
%
TOFFOLI 0,1,9
TOFFOLI 9,3,2
TOFFOLI 0,1,9
%
TOFFOLI 0,3,1
CNOT 3,0
%
H 3
CR 2,3,"pi/2"
CR 1,3,"pi/4"
CR 0,3,"pi/8"
H 2
CR 1,2,"pi/2"
CR 0,2,"pi/4"
H 1
CR 0,1,"pi/2"
H 0
SWAP 0,3
SWAP 1,2
%
DAGGER
H 3
CNOT 3,2
CNOT 3,1
CNOT 3,0
%
TOFFOLI 0,1,9
TOFFOLI 9,3,2
TOFFOLI 0,1,9
%
TOFFOLI 0,3,1
CNOT 3,0
ENDDAGGER
%%%%%%%%%%%%%%%
%RY
%RY0%%%%%%%%%%%%
X 2
X 1
%
TOFFOLI 0,2,4
TOFFOLI 0,2,5
RY 4,"-pi/4"
RY 5,"-pi/4"
TOFFOLI 0,2,4
TOFFOLI 0,2,5
RY 4,"pi/4"
RY 5,"pi/4"
%
TOFFOLI 0,1,4
TOFFOLI 0,1,5
RY 4,"-pi/8"
RY 5,"-pi/8"
TOFFOLI 0,1,4
TOFFOLI 0,1,5
RY 4,"pi/8"
RY 5,"pi/8"
%
CNOT 0,4
CNOT 0,5
RY 4,"-pi/16"
RY 5,"-pi/16"
CNOT 0,4
CNOT 0,5
RY 4,"pi/16"
RY 5,"pi/16"
%
X 2
X 1
```

```
%%%%%%%%%%%%%%%%
X 2
CNOT 2,1
%%%
TOFFOLI 0,1,6
TOFFOLI 0,1,7
RY 6,"-pi/4"
RY 7,"-pi/4"
TOFFOLI 0,1,6
TOFFOLI 0,1,7
RY 6,"pi/4"
RY 7,"pi/4"
%
CNOT 0,6
CNOT 0,7
RY 6,"-pi/8"
RY 7,"-pi/8"
CNOT 0,6
CNOT 0,7
RY 6,"pi/8"
RY 7,"pi/8"
%
CNOT 2,1
X 2
%RY1%%%%%%%%%%%%
X 0
%
TOFFOLI 0,1,4
TOFFOLI 0,1,5
RY 4,"-pi/6"
RY 5,"-pi/6"
TOFFOLI 0,1,4
TOFFOLI 0,1,5
RY 4,"pi/6"
RY 5,"pi/6"
%
TOFFOLI 0,1,9
X 2
%
TOFFOLI 9,2,6
TOFFOLI 9,2,7
RY 6,"-pi/4"
RY 7,"-pi/4"
TOFFOLI 9,2,6
TOFFOLI 9,2,7
RY 6,"pi/4"
RY 7,"pi/4"
%
CNOT 9,6
CNOT 9,7
RY 6,"-pi/8"
RY 7,"-pi/8"
CNOT 9,6
CNOT 9,7
RY 6,"pi/8"
RY 7,"pi/8"

X 2
TOFFOLI 0,1,9
%
X 0
%
%Anc
%
TOFFOLI 4,5,10
TOFFOLI 6,7,11
TOFFOLI 10,11,8
%CB+
DAGGER
%
X 3
%%%%%%%%%%%%%%%%%
H 3
CNOT 3,2
CNOT 3,1
CNOT 3,0
%
TOFFOLI 0,1,9
TOFFOLI 9,3,2
TOFFOLI 0,1,9
%
TOFFOLI 0,3,1
CNOT 3,0
%
H 3
CR 2,3,"pi/2"
CR 1,3,"pi/4"
CR 0,3,"pi/8"
```

```
H 2
CR 1,2,"pi/2"
CR 0,2,"pi/4"
H 1
CR 0,1,"pi/2"
H 0
SWAP 0,3
SWAP 1,2
%
DAGGER
H 3
CNOT 3,2
CNOT 3,1
CNOT 3,0
%
TOFFOLI 0,1,9
TOFFOLI 9,3,2
TOFFOLI 0,1,9
%
TOFFOLI 0,3,1
CNOT 3,0
ENDDAGGER
%%%%%%%%%%%%%%%
ENDDAGGER
%
PMEASURE 8,2,1,0
```

## 2. Realization code for *n*=2

## The code for ibmq_santiago

```
from qiskit import QuantumRegister, ClassicalRegister, QuantumCircuit, execute
from qiskit.visualization import plot_histogram
from numpy import pi
import math

from qiskit import IBMQ
provider = IBMQ.load_account()

backend = provider.get_backend('ibmq_santiago')

qreg_q = QuantumRegister(5, 'q')
creg_c = ClassicalRegister(3, 'c')
circuit = QuantumCircuit(qreg_q, creg_c)
#init b
circuit.u(pi/2, 0, 0, qreg_q[1])
circuit.cx(qreg_q[1], qreg_q[2])
circuit.u(pi, 0, pi, qreg_q[1])
circuit.u(pi/4, 0, 0, qreg_q[2])
circuit.cx(qreg_q[1], qreg_q[2])
#CB
circuit.u(0, 0, pi/2, qreg_q[1])
circuit.u(pi/2, pi/4, -pi/2, qreg_q[2])
circuit.cx(qreg_q[2], qreg_q[1])
circuit.u(2*pi/3, 1.195913276*pi, -0.195913276*pi, qreg_q[1])
circuit.u(4*pi/3, -0.195913276*pi, 1.195913276*pi, qreg_q[2])
circuit.cx(qreg_q[2], qreg_q[1])
#cRy
circuit.u(0.27877348*pi, 0, 0, qreg_q[0])
circuit.u(pi, 0, pi/2, qreg_q[1])
circuit.u(pi/2, 0, pi, qreg_q[2])
circuit.cx(qreg_q[1], qreg_q[0])
circuit.u(0, 0, -pi/2, qreg_q[2])
circuit.u(-0.27877348*pi, 0, 0, qreg_q[0])
circuit.cx(qreg_q[1], qreg_q[0])
#SWAP
circuit.cx(qreg_q[2], qreg_q[1])
circuit.cx(qreg_q[1], qreg_q[2])
circuit.cx(qreg_q[2], qreg_q[1])
#cRy
circuit.u(0.046783656*pi, 0, 0, qreg_q[0])
circuit.cx(qreg_q[1], qreg_q[0])
circuit.u(-0.046783656*pi, 0, 0, qreg_q[0])
circuit.cx(qreg_q[1], qreg_q[0])
circuit.u(0.080430623*pi, 0, 0, qreg_q[0])
circuit.u(pi, 0, pi, qreg_q[1])
circuit.cx(qreg_q[1], qreg_q[0])
circuit.u(-0.080430623*pi, 0, 0, qreg_q[0])
circuit.cx(qreg_q[1], qreg_q[0])
#CB+
circuit.u(pi, 0, pi, qreg_q[2])
circuit.u(pi/2, pi, 3*pi/2, qreg_q[1])
```

```
circuit.u(0, 0, -pi/2, qreg_q[2])
circuit.cx(qreg_q[1], qreg_q[2])
circuit.u(4*pi/3, 0.195913276*pi, -1.195913276*pi, qreg_q[1])
circuit.u(2*pi/3, -1.195913276*pi, 0.195913276*pi, qreg_q[2])
circuit.cx(qreg_q[1], qreg_q[2])
circuit.u(pi/2, pi/2, pi, qreg_q[1])
circuit.u(0, 0, pi/2, qreg_q[2])
circuit.measure(qreg_q[0], creg_c[2])
circuit.measure(qreg_q[2], creg_c[0])
circuit.measure(qreg_q[1], creg_c[1])
job = execute(circuit, backend, initial_layout=[2,3,4,0,1], optimization_level=0, shots=4096)
result = job.result()
counts = result.get_counts(circuit)
circuit.draw('mpl').show()
plot_histogram(counts).show()
```